\documentclass{Interspeech2024}

\interspeechcameraready

\title{Speculative Speech Recognition\\by Audio-Prefixed Low-Rank Adaptation of Language Models}

\name[affiliation={1,2,3}]{Bolaji}{Yusuf}
\name[affiliation={1}]{Murali}{Karthick Baskar}
\name[affiliation={1}]{Andrew}{Rosenberg}
\name[affiliation={1}]{Bhuvana}{Ramabhadran}

\address{
  $^1$Google Inc., USA \\
  $^2$Bogazici University, Turkey \\
  $^3$Brno University of Technology, Speech@FIT, Czechia
  }
\email{\{iyusuf\}@fit.vut.cz, \{mkbaskar,rosenberg,bhuv\}@google.com}

\usepackage{amsmath}

\usepackage{booktabs}
\usepackage{cite}
\usepackage{mathtools}
\usepackage{pgfplots}

\newcommand{\Matrix}[1]{\mathbf{\textbf{#1}}}

\DeclareMathOperator*{\argmin}{arg\,min}

\newcommand{\vsxspace}[1]{}
\expandafter\def\expandafter\normalsize\expandafter{%
    \normalsize%
    \setlength\abovedisplayskip{0pt}%
    \setlength\belowdisplayskip{0pt}%
    \setlength\abovedisplayshortskip{-7pt}%
    \setlength\belowdisplayshortskip{2pt}%
}

\pgfplotsset{compat=1.16}

\begin{document}
\maketitle
 
\begin{abstract}
This paper explores speculative speech recognition (SSR), where we empower conventional automatic speech recognition (ASR) with speculation capabilities, allowing the recognizer to run ahead of audio.
We introduce a metric for measuring SSR performance and
we propose a model which does SSR by combining a RNN-Transducer-based ASR system with an audio-prefixed language model (LM).
The ASR system transcribes ongoing audio and feeds the resulting transcripts, along with an audio-dependent prefix, to the LM, which speculates likely completions for the transcriptions.
We experiment with a variety of ASR datasets on which show the efficacy our method and the feasibility of SSR as a method of reducing ASR latency.

\end{abstract}

\section{Introduction}
The experience of users interacting with an automatic speech recognition (ASR) system is colored by its latency---how quickly it is able to respond to user requests---in addition to its accuracy.
A system which can respond quickly to user queries is generally preferred to a slower one with similar accuracy.

There has therefore being considerable effort towards improving ASR latency, such as using lightweight, fully-causal or limited-context encoders~\cite{he2019streaming,sainath19_interspeech,zhang2020transformer,strimel2023lookahead}, and using modified training objectives such as timing penalties~\cite{li2020towards} and FastEmit~\cite{yu2021fastemit} which encourage early output symbol emission to counteract the tendency of limited-context models to delay emission until enough context has been accumulated to make a confident decision.
These methods aim to make the latency as close to zero as possible without incurring significant degradation recognition accuracy, i.e.,
the best case for these approaches is that the model finishes transcription just as the user finishes speaking.
However, ASR is only the first step in user interaction and it is often followed by some form of natural language processing (NLP) such as information retrieval or machine translation or spoken language understanding.
Therefore, even if ASR latency were to reach zero, the overall end-to-end latency experienced by the user would still above zero.

Prefetching~\cite{chang2020low,li2021better} provides a template for reducing the end-to-end latency further.
The method hinges on the observation that there is a delay between the an ASR system's emission of the last output symbol and being able to confidently determine that the utterance has ended.
ASR hypotheses are sent downstream as soon as a token is emitted without awaiting the end-of-utterance confirmation, and the downstream computation commences immediately.
Thus, the endpointing latency is mitigated in exchange for extra computational overhead.

In this paper, we tackle speculative speech recognition (SSR)%
\footnote{Note that this bears no direct relation to speculative decoding in language modelling~\cite{leviathan2023fast,chen2023accelerating}, which involves using a small language model to speed up inference in a larger model.}%
, the problem of accurately generating the \emph{full} transcription \emph{before} the user has finished speaking.
Being able to solve this problem would allow any downstream NLP operations to be initiated earlier, and therefore further reduce end-to-end latency.
Conceptually, this problem has two parts: transcription and speculation,
where the former corresponds to the generation of textual tokens whose corresponding speech has actually been uttered and the latter corresponds to the generation of tokens which have no corresponding input speech.

Schwarz et. al.~\cite{schwarz23_interspeech} recently proposed a method for SSR.
Their approach---which we take as baseline in this work---uses a hybrid of an RNN-Transducer (RNN-T)~\cite{graves2012sequence} ASR system and a pretrained language model (LM), where the RNN-T transcribes the incomplete spoken utterance and the resulting hypotheses are fed into the LM to speculate likely completions.

The ASR-LM hybrid is however limited in that the LM is pretrained for generic text completion, and, consequently, does not account for the idiosyncrasies of operating on ASR output.
Specifically, it doesn't account for the fact that its input is a \emph{hypothesis} from an ASR system and may thus contain errors; it also does not consider information contained in the audio signal which may be lost in the transcript, but would nevertheless be useful for speculation.
Therefore, we propose a modified scheme which prepends an audio-dependent soft prompt~\cite{lester-etal-2021-power} to the ASR hypothesis as input to the LM.
We further finetune the LM with low-rank (LoRA) adapters~\cite{hu2022lora} which are trained---along with the soft prompt---to speculate the ideal suffix tokens to complete the ASR hypothesis,
thereby allowing the model to have knowledge of both its acoustic context and the possibility of errors from ASR when making decisions.

As the main contributions of this work, we propose:
\begin{itemize}
    \item A Conformer-Transformer hybrid model which uses an audio-conditioned prefix LM for SSR, with experiments showing the efficacy of the proposed system on public datasets, and its superiority to purely LM-based speculation.
    \item An edit-distance-based alignment procedure to get training labels for error-aware speculative ASR systems.
    \item Suffix Oracle Word Error Rate (SOWER), a metric for measuring the performance of a speculative ASR system, which accounts for various peculiarities of the problem.
\end{itemize}

\section{Methods}
\begin{figure}
    \centering
    \includegraphics[width=0.55\linewidth]{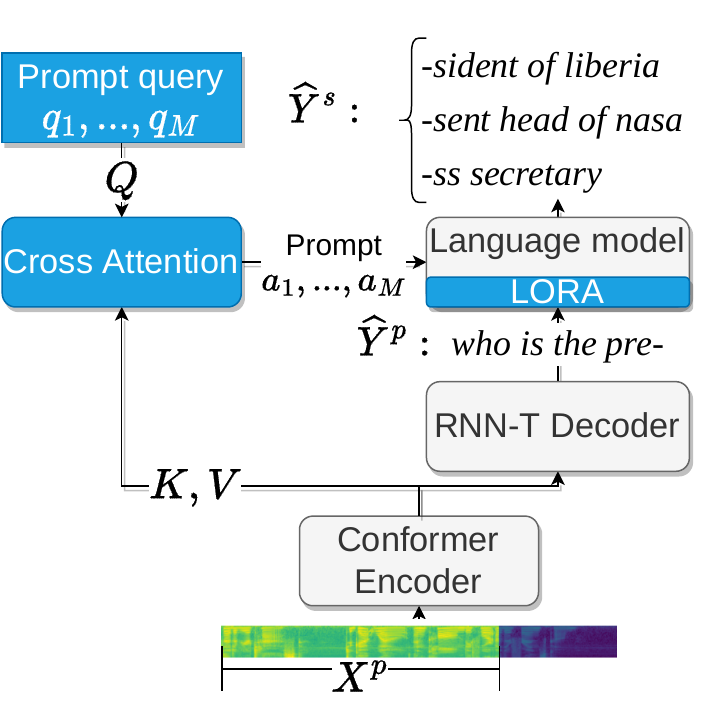}
    \caption{Illustration of the proposed model with trainable parameters in blue. A Conformer-Transducer ASR model decodes the speech into text. Then a language model is prompted with a prefix computed from the Conformer encoder output to predict likely completions for the partial ASR hypothesis.}
    \label{fig:flowchart}
    \vspace{-4.1mm}
\end{figure}
In conventional ASR, the goal is to predict a sequence of tokens $Y=(y_1, y_2, \dots, y_U)$ given a sequence of acoustic inputs $\Matrix{X} = (\Matrix{x}_1, \Matrix{x}_2, \dots, \Matrix{x}_T)$.
This generally involves modelling the conditional probability:
\vsxspace{-12pt}
\begin{align}
P(Y|\Matrix{X}) = \prod_{u=1}^U P(y_u | \Matrix{X}, y_{< u} ).
\end{align}
In speculative ASR, instead of the full acoustic input ($\Matrix{X}$), the input is a prefix comprising its first $j$ frames, $\Matrix{X}^p = (\Matrix{x}_1, \Matrix{x}_2, \dots, \Matrix{x}_j): j < T$, and the goal is to model the \emph{full} output sequence, $Y$, as if the input were the entirety of $\Matrix{X}$:
\vsxspace{-8pt}
\begin{align}
P(Y|\Matrix{X}^p) = \prod_{u=1}^U P(y_u | \Matrix{X}^p, y_{< u} ).
\label{eqn:speculative_asr}
\end{align}

\subsection{Baseline speculative ASR with ASR-LM hybrid}
\label{sec:method:spec}
Our approach to solving speculative ASR is based on the hybrid approach from~\cite{schwarz23_interspeech}, with the LSTMs in the RNN-T and LM replaced by a pretrained Conformer-Transducer~\cite{gulati20interspeech} ASR model and a pretrained Transformer language model~\cite{vaswani2017attention}.
The Transducer is used to transcribe the spoken prefix $\Matrix{X}^p$ into ${Y}^p = ({y}_1, \dots, {y}_r)$, and the LM is used to predict the suffix $Y^s = (y_{r+1} \dots, y_U)$ by conditioning on the transcription.
This essentially parameterizes~\eqref{eqn:speculative_asr} thus:
\vsxspace{-10pt}
\begin{align}
    P(Y|\Matrix{X}^p) &= \underbrace{\prod_{u=1}^r P_{\phi}(y_u | \Matrix{X}^p, y_{< u} )}_{\text{Transcription by RNN-T}}
    \cdot \underbrace{ \prod_{u=r+1}^U P_{\theta}(y_u | y_{< u} )}_{\text{Speculation by LM}},
    \label{eqn:catlm}
\end{align}
where $\phi$ denotes the parameters of the Transducer model and $\theta$ denotes the parameters of the language model.

\subsection{Audio-aware speculative ASR}
\label{sec:method:audio_spec}
Simply stacking the Transducer and the LM as in Section~\ref{sec:method:spec} implicitly assumes that the transcription is an efficient enough summary of the input speech for the purposes of speculating the suffix, i.e., that $P(Y^s|X^p,Y^p) = P(Y^s|Y^p)$.
However, the speech signal itself contains information such as speaker, channel or domain clues which could be useful for better constraining the language model output.
Furthermore, since the Transducer objective considers all possible alignments
(in contrast to hybrid models which use forced alignment)
,
there is only a loose correspondence between the input frame and the output token ($j$ and $r$ respectively in Section~\ref{sec:method:spec}), meaning that the ASR can choose to transcribe sounds later than they are uttered.
We argue, therefore, that it would be useful to condition the suffix generation on a representation of the audio signal.

To this end, we add a fixed-length representation,  $\Matrix{A}^p = (\Matrix{a}_1, \dots, \Matrix{a}_M)$, of the audio as a prefix to the language model.
$\Matrix{A}^p$ is the output of a multihead attention layer whose keys and values are projections of the Transducer's Conformer encoder output, and whose queries are projections a sequence of $M$ trainable vectors, $\Matrix{Q}=(\Matrix{q}_1,\dots, \Matrix{q}_M)$.
Thus, when the Transformer speculates the $u$th overall output token, its input is:
\vsxspace{-4pt}
\begin{align}
c_u = \bigl(\Matrix{a}_1, \Matrix{a}_2, \dots, \Matrix{a}_M, \Matrix{e}[y_1], \Matrix{e}[y_2], \dots \Matrix{e}[y_{u-1}]\bigr),
\end{align}
where $\Matrix{e}[\cdot]$ denotes the LM's input embedding lookup.
Although it is possible in theory to use other speech representations, we choose the Conformer output since we get it for free as part of the ASR computation. 
Moreover, we know that it is rich in lexical content since it is an encoding that is used directly for ASR.
We use a fixed-length summary of the encoding as prefix because doing so reduces (when $j>M$), and keeps fixed, the computational cost of the subsequent LM decoding compared to using the encoding directly.
Ultimately, instead of~\eqref{eqn:catlm}, we have the following parameterization:
\vsxspace{-8pt}
\begin{align}
    P(Y|\Matrix{X}^p) = P_{\phi}(Y^p|\Matrix{X}^p)
    \prod_{u=r+1}^U P_{\phi_e, \theta, Q, \zeta}(y_u | y_{< u}, \Matrix{X}^p ),
    \label{eqn:speech_prefix}
\end{align}
where $\zeta$ denotes the multihead attention parameters.
Our training process will involve keeping the first term on the right hand side fixed, and learning parameters which maximize the second term---$P_{\phi_e, \theta, Q, \zeta}(Y^s | Y^p, \Matrix{X}^p ) \coloneqq \prod_{u=r+1}^U P_{\phi_e, \theta, Q, \zeta}(y_u | y_{< u}, \Matrix{X}^p )$.

\subsection{Alignment and finetuning for speculation}
\label{sec:method:training}
With the parameters of the Conformer-Transducer fixed, we add LoRA layers to each Transformer layer in the LM.
The LoRA parameters are trained along with the LM's tied embedding-softmax layer, the cross-attention parameters and query vectors to maximize the log-likelihood of the correct suffix.

We do the finetuning on a dataset of audio-text transcription $(\Matrix{X}, Y)$ pairs.
For each training sample, we first get $\Matrix{X}^p$ by truncating the last \textit{1 second} of audio.
Then we pass the truncated audio to the Transducer-based ASR system to get a hypothesized transcript $\hat{Y}^p$.
Next we feed this hypothesis into the language model---along with the prefix from the cross-attention on the encoder output---to predict $Y^s = (y_{r+1}, \dots, y_U)$---the portion of $Y$ corresponding to the truncated 1 second of audio.

Determining $r$ for training, however, is not as trivial as it may seem.
Consider, for example, an utterance whose correct transcription is \textit{``i'd like to call my father"}, but for which the ASR generates the prefix \textit{``i'd line to call ma-"}.
Because of the ASR errors, the correct suffix is not clearly defined.
It is therefore necessary to first determine what part of the transcription has been covered---possibly erroneously---by the ASR and what part remains to be speculated.

\begin{figure}
    \centering
    \includegraphics[width=0.7\linewidth]{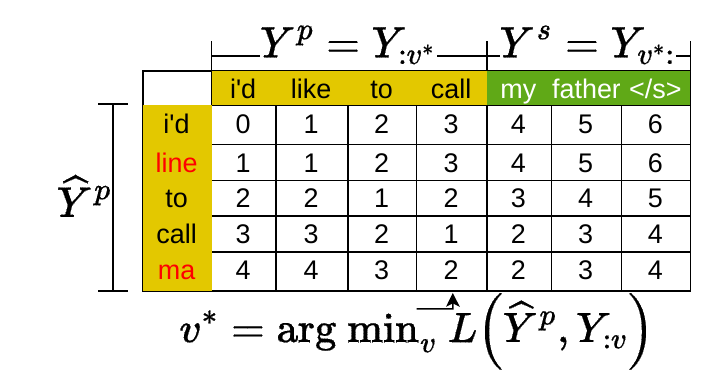}
    \caption{AWSED procedure for computing the optimal alignment between a hypothesis prefix and a full reference.
    }
    \label{fig:awsed}
\end{figure}

\noindent\textbf{AWSED}: We propose solving this problem by Alignment With Subsequence Edit Distance (AWSED).
This procedure, illustrated in Figure~\ref{fig:awsed}, involves computing the Levenshtein distance (LVD)~\cite{levenshtein1966binary} between $\hat{Y}^p$ and all left-substrings of $Y$.
More concretely, taking the left- and right-substring of $Y$ at some index $v$ respectively as $Y_{:v} \coloneqq (y_1, \dots, y_v)$ and $Y_{v:} \coloneqq (y_{v+1}, \dots, y_U)$, then the desired target suffix, $Y^s = Y_{v^*:}$, where:
\begin{align}
    v^* = \argmin_v \ L(\hat{Y}^p, Y_{:v}),
\end{align}
where $L(s_1, s_2)$ is the LVD between the strings $s_1$ and $s_2$.
Note that rather than having to compute $L(\hat{Y}^p, Y_{:v})$ separately for each $v$, we need only one run of the dynamic programming algorithm for computing the LVD to get $v^*$
because the last row of the matrix of accumulated costs already contains $L(\hat{Y}^p, Y_{:v})$ for every $v$, and we only need to take the $\argmin$ on this row.
In case multiple indices are tied for the $\argmin$, we pick the leftmost one.
In our running example, the AWSED procedure yields the resulting $Y^s$ as \textit{``-my father"}.
Had we taken the rightmost $\argmin$, then $Y^s$ would have been \textit{``-father"}.
This choice however does not affect the overall word edit distance from the correct transcription, which is 2 in both cases (1 substitution and 1 insertion for \textit{``i'd {line} to call {ma} my father"}, and 2 substitutions for \textit{``i'd {line} to call {ma} father"}).

We use stochastic gradient descent with Adam~\cite{DBLP:journals/corr/KingmaB14} to finetune. For each mini-batch $\mathcal{B}$, we minimize the cross-entropy:
\vsxspace{-18pt}
\begin{align}
    J_{\mathcal{B}} = -\sum_{ (\Matrix{X}, Y) \in \mathcal{B} } \sum_{u={v^* +1}}^U \log P_{\phi_e, \theta + \Delta \theta, \Matrix{Q}, \zeta}(y_u | \hat{Y}^p, \Matrix{X}^p ),
\end{align}
with respect to the cross-attention parameters ($\zeta$), soft prompt query vectors ($\Matrix{Q}$) and LoRA parameters ($\Delta \theta$).

\section{Experiments}
\subsection{Metrics}
SSR differs from conventional ASR in one crucial respect: it has no single correct answer; for a given prefix audio, it is impossible to determine the suffix with perfect certitude.
This makes the word error rate (WER) metric
unsuitable for out purposes.
Moreover WER---or any other metric that is computed over the entire sentence---also encompasses a measure of the prefix transcription accuracy and thus can only be, at best, a diffuse proxy for speculation performance.
Hence, we seek a metric which isolates the accuracy of the speculated suffix while also accounting for the uncertainty inherent in the task.

One solution would be to treat the task as a language modeling one, and to report the perplexity of the correct suffix.
Perplexity however has the drawback that it is not an operational measure, i.e., we can use it to compare systems and say lower is better, but knowing the exact perplexity score says little about the conditions under which the system is usable.
Furthermore, perplexity has to be computed with ``teacher-forcing", which does not reflect the practical usage of an ASR system.

Instead, we treat speculation as a quasi-retrieval problem and propose an analog of recall-at-k.
The metric, which we term suffix oracle word error rate (SOWER) is computed by truncating $t$ seconds from the end of the audio, letting the system hypothesize $k$ suffixes $\hat{Y}^s(1), \dots, \hat{Y}^s(k)$, and returning the minimal WER between the hypotheses and the correct suffix $Y^s$:
\vsxspace{-15pt}
\begin{align}
    S(t, k) = \min_{i\in [1, k]} \text{WER} (\hat{Y}^s(i), Y^s).
\end{align}
This measures how well we expect the model to do if we return the top-$k$ hypotheses.
By default, we set $t=1$ and $k=8$ in our experiments.
Note that, to get the correct suffix for evaluation, we once again use the AWSED procedure described in Section~\ref{sec:method:training} on the prefix transcription and the full reference.

\subsection{Datasets and model architecture}
We use Speechstew~\cite{chan2021speechstew}, an amalgamation of multiple public corpora totalling about 5000 hours, to pretrain the Conformer for transcription.
We conduct two sets of experiments, varying the pretraining, finetuinnig and testing data:

\noindent\textbf{Librispeech-only}: Here, we use Librispeech LM data \cite{panayotov2015librispeech} for pretraining the language model, finetune on the 960h Librispeech training set, and test on the Librispeech test splits.

\noindent\textbf{Multi-domain}: Here, we pretrain the LM with data composed of the Librispeech LM data along with text from the Switchboard~\cite{godfrey1992switchboard}, TED-LIUM~\cite{rousseau2012ted,hernandez2018ted} and Wall Street Journal~\cite{paul-baker-1992-design} corpora. We finetune on Speechstew, and test on the AMI-IHM~\cite{kraaij2005ami}, Librispeech, Switchboard and TED-LIUM test sets.

The Transducer model has a 100 million parameter, 17-layer Conformer-L~\cite{gulati20interspeech} encoder with 512-dimensional layers operating on 80-dimensional log-mel filterbank inputs, an LSTM prediction network with one 512-dimensional layer, and a 2-layer feedforward joint network with 512-dimensional intermediate layer and 1024-dimensional output softmax layer corresponding to 1024 Librispeech word-piece~\cite{kudo-richardson-2018-sentencepiece} targets.

The LM is a 100 million parameter transformer decoder~\cite{vaswani2017attention} with eight 1024-dimensional layers, each split into 16 attention heads, and tied embedding-softmax with the same 1024 word piece units as the Transducer.

We use $M=64$ queries (equivalent in length to $2.56$ seconds of audio) of 1024 dimensions and a single 1024-dimensional cross-attention layer with four heads for computing the soft-prompt.
We set the rank of the LoRA adapters to 10 for the Librispeech experiments and 50 for the multi-domain experiments, resulting in 12 million and 19 million trainable parameters respectively to be finetuned for speculation.

\subsection{Speculation systems}
We report results for four model configurations:

\noindent \textbf{Pretrained model (PM)}: This is the our replication of the model from~\cite{schwarz23_interspeech} (described in \S\ref{sec:method:spec}), which uses the pretrained LM to speculate suffixes without any finetuning or audio prefix.
We note that~\cite{schwarz23_interspeech} also incorporates a confidence model on top of speculation.
Here, we focus only on the speculator itself.

\noindent \textbf{Hypothesis-only finetuned (HO)}: Similar to PM, this configuration only uses the Transducer hypotheses---without audio encoding---as the LM input.
The LM, however, is finetuned for speculation.
Instead of the audio-aware prefix, $\Matrix{A}^p$, we use the trainable vectors, $\Matrix{q}_1, \dots, \Matrix{q}_M$, directly as the (audio-agnostic) prefix and finetune them along with the LM softmax and LoRA layers.
This configuration allows us to separate what improvements, if any, come from ASR-error-aware finetuning and what improvements come from using an audio prefix.

\noindent \textbf{Speech-prefix finetuned (SP)}: This is the full configuration as described in \S\ref{sec:method:audio_spec}, which uses an audio-dependent soft prompt.

\noindent \textbf{Speech prefix + Text Injection (ST)}: This configuration adds text injection~\cite{wang21t_interspeech,yusuf22usted} to SP.
While finetuning on paired speech-text data, the trainable parameters are jointly trained on the text used to pretrain the respective LM, so that the LM does not overfit to the paired training data.
Since the text-only training mini-batches have no audio input, their cross-attention keys and values and, consequently the LM prompts, are set to 0.
\begin{table}[t]
    \centering
    \caption{SOWER on the Librispeech dev and test sets; ``tavg" is the average of the test sets.
    PM-1w and ST-1w denote SOWER computed on just the first suffix word;
    $\Delta_{97.5}$ and $\Delta_{2.5}$ denote respectively the $97.5$ and $2.5$ percentile values of the improvement over PM estimated with blockwise bootstrap~\cite{bisani2004bootstrap,liu20c_interspeech}.
    }
    \resizebox{!}{0.168\linewidth}{
    \begin{tabular}{lcccc|ccc}
    \toprule
         & dev-c & dev-o & test-c & test-o &  tavg&$\Delta_{97.5}$&$\Delta_{2.5}$\\
         \midrule
         PM &  75.0 &	79.4&	74.4&	81.2 & 77.8 &0 &0\\
         HO &  69.0&	73.2	& 69.0 &	75.0 & 72.0 & 5.6 & 6.2 \\
         SP &  64.1 &	68.7&	64.8	&69.6 & 67.2&9.9 & 11.4\\
         ST & 61.0&	66.6&	61.7	&66.9 &64.3 & 12.8 & 14.2\\
         \midrule
         PM-1w & 61.3&	67.2&	62.0&	68.1 & 65.1& 0 & 0\\
         ST-1w & 46.1&	51.7&	46.4&	52.1& 49.3 & 15.3 & 15.8\\
         \bottomrule
    \end{tabular}
    }
    \label{tab:ls_s_100_8}
\end{table}

\begin{table}[t]
    \centering
    \caption{Oracle WER on the Librispeech dev and test sets. WERR denotes the percentage tavg WER recovered by speculation, computed as $100*\frac{sys-baseline}{topline - baseline}$, and ST(k=1) refers to using 1-best speculation from ST instead of 8-best.
    }
    \resizebox{!}{0.2\linewidth}{
    \begin{tabular}{lcccc|cc}
    \toprule
         & dev-c & dev-o & test-c & test-o & tavg & WERR \\
         \midrule
         Baseline & 11.3	& 16.3 &	11.6	&15.6 & 13.6 & 0\\
         \midrule
         PM &  8.9	& 13.9	&9.1 &	13.6 & 11.4 & 21.8 \\
         HO &  8.3 &	13.1 &	8.5 &	12.7 & 10.6 & 29.7 \\
         SP &  7.8 &	12.5 &	8.1&	12.1 & 10.1 & 34.7\\
         ST & 7.5 &	12.3 &	7.8 &	11.8 & 9.8 & 37.6\\
         \midrule
         {ST (k=1)} & 10.2&	15.3&	10.4&	14.6 & 12.5 & 10.9\\
         \midrule
         Topline & 2.1 &	4.7&	2.2&	4.8 & 3.5 & 100 \\
         \bottomrule
    \end{tabular}
    }
    \label{tab:ls_w_100_8}
    \vspace{-3mm}
\end{table}

\subsection{Librispeech results}
First we conduct experiments of the Librispeech test sets, where we truncate the last second of audio from each utterance and speculate them with various systems.
This 1 second of audio contains an average of around 2 words/utterance (1.98-2.2 depending on the test set) which the systems must speculate.

Table~\ref{tab:ls_s_100_8} shows SOWER---$S(1,8)$---on the Librispeech test sets. All the systems, including PM, yield SOWER below 100, i.e., on average, picking the best of top-8 speculated hypotheses is better than not speculating at all.
HO significantly outperforms PM, highlighting the positive impact of making the LM ASR-error-aware.
SP yields a further 6.7\% relative improvement over HO.
Finetuning jointly with unpaired text (ST) leads to further improvements, indicating that the SP---and HO---loses some capacity as a language model while fitting the ASR training set (which is much smaller than the unpaired data), and this capacity can be somewhat recovered by joint training.
Finally, the table also shows that speculating only the next word is, as expected, easier than predicting the rest of the utterance---with ST-1w in particular averaging SOWER below 50\% on single word prediction.

On inspecting the transcriptions and listening to the corresponding audio, we found several cases where the prefix audio ends with the beginnings of a sound, usually a stop consonant, which the ASR does not transcribe.
Where the systems without audio prompt speculate semantically-appropriate completions, the ones with audio prompt speculate semantically-appropriate completions which also start with the correct phoneme.

Table~\ref{tab:ls_w_100_8} shows the Librispeech oracle WER computed over the entire utterances (not just the suffixes), and therefore show the impact of speculation on the whole WER.
The speculation methods are compared with two purely Conformer-Transducer systems: a baseline, where each utterance is truncated and no speculation is done, and a topline which sees the entire audio without truncation or need to speculate.
Overall, the oracle word error rates of the speculation methods follows the same trends as the SOWER, with the best speculation method (ST) recovering about 37.6\% of the gap between the baseline and the topline.

Table~\ref{tab:ls_w_100_8} also shows the oracle WER of the best system, ST, at $k=1$.
Unsurprisingly, the oracle WER (and SOWER although it is not shown) degrades as $k$ is decreased.
More interestingly, we see that even at $k=1$, ST outperforms the baseline.
In other words, even one-shot speculation with ST is slightly better than no speculation at all.
In fact, we found that all the speculation systems improve upon the baseline with the exception of PM at $k=1$ (which has SOWER $>100$).

\begin{table}[t]
    \centering
    \caption{SOWER on various test sets.}
    \resizebox{!}{0.125\linewidth}{
    \begin{tabular}{lccccc|ccc}
    \toprule
         & ami & test-c & test-o & swbd &ted& tavg&$\Delta_{97.5}$&$\Delta_{2.5}$\\
         \midrule
         PM &  95.6&	80.0	&84.2&	93.7&	89.8 &88.7 &0 & 0\\
         HO &  88.6&	80.2&	84.1&	89.8&	86.3&85.8 & 1.4 & 5.8 \\
         SP &  79.2&	69.7&	73.7&	82.3&	76.1&76.2 & 10.4 & 15.1\\
         ST & 79.6&	67.0&	71.6&	83.2&	77.8&75.8 & 10.2 & 15.0\\
         \bottomrule
    \end{tabular}
    }
    \label{tab:all_s_100_8}
\vspace{-5mm}
\end{table}

\subsection{Multi-domain results}
In addition to Librispeech, we report results on the AMI, Switchboard and TED-LIUM test sets in order to test generalization of the proposed speculation methods across domains.

Table~\ref{tab:all_s_100_8} shows the SOWER across test sets.
Here we find that the improvements from HO compared to PM are more subdued.
In fact, for the Librispeech test sets, HO is slightly worse than PM.
The largest improvements are on AMI, which is not represented in the LM pretraining data, and thus benefits the most from finetuning on the Speechstew training set--which does contain AMI data.
SP outperforms HO by a larger margin across all datasets, underscoring the importance of conditioning the LM on audio.
Finally, we observe that ST does not outperform SP in terms of SOWER, except on Librispeech.
This is likely due to the fact that the Librispeech text is the most represented in the unpaired text used for joint training.

Qualitatively, we found that because the Speechstew data is an amalgamation of various datasets, some of which have transcripts with punctuation and extra tags like ``[laughter]", ``[noise]" etc., HO learns to frequently speculate these tags even for datasets like Librispeech which are normalized, and thus increases number of errors, while SP does so less frequently.
This explains some of the degradation of HO on Librispeech.
It also hints at another advantage of having audio conditioning, namely it triggers domain-specific behavior in the LM.


\section{Conclusions}
In this paper, we've tackled the problem of doing ASR before the getting the input by using an LM to speculate the missing parts.
We propose an approach to reducing the entropy in speculation by feeding a fixed-length audio prefix to the LM, and a mechanism for finetuning the LM in the presence ASR-errors.
 This speculative model is able to correctly retrieve 35.7\% of suffixes up to one second ahead of time---and 50.7\% when predicting only the next word, showing its viability as a means of achieving negative latency in ASR.

Future work in this direction include finding other ways of reducing the suffix entropy such as incorporating user-specific language models (as done in~\cite{schwarz23_interspeech}) or biasing lists, retrieving from related documents~\cite{chan2023using,yusuf2023fly,wu23e_interspeech} or simply using better LMs.
Furthermore, although we show that we can run ahead-of-audio, fully realizing the latency improvements would require using efficient LMs and LM inference schemes such as~\cite{leviathan2023fast,chen2023accelerating} especially for larger LMs.
Another interesting direction is to incorporate speculation directly into the ASR---the transducer objective, for instance, places no limits on input and output lengths.

\bibliographystyle{IEEEtran}
\bibliography{mybib}

\end{document}